\documentclass[manuscript]{aastex}

\slugcomment{To appear in PASP}

\begin{document}

\title{Temporal Variations of Telluric Water Vapor Absorption at Apache Point Observatory}

\author{Dan Li\altaffilmark{1}, Cullen H. Blake\altaffilmark{1}, David Nidever\altaffilmark{2}, and Samuel Halverson\altaffilmark{1,3}}

\altaffiltext{1}{Department of Physics \& Astronomy, University of Pennsylvania, 209 South 33rd Street, Philadelphia, PA 19104, USA}
\altaffiltext{2}{National Optical Astronomy Observatory, 950 North Cherry Ave, Tucson, AZ 85719, USA}
\altaffiltext{3}{Sagan Fellow}

\begin{abstract}
Time-variable absorption by water vapor in Earth's atmosphere presents an important source of systematic error for a wide range of ground-based astronomical measurements, particularly at near-infrared wavelengths. We present results from the first study on the temporal and spatial variability of water vapor absorption at Apache Point Observatory (APO). We analyze $\sim$400,000 high-resolution, near-infrared ($H$-band) spectra of hot stars collected as calibration data for the APO Galactic Evolution Experiment (APOGEE) survey. We fit for the optical depths of telluric water vapor absorption features in APOGEE spectra and convert these optical depths to Precipitable Water Vapor (PWV) using contemporaneous data from a GPS-based PWV monitoring station at APO. Based on simultaneous measurements obtained over a 3$^{\circ}$ field of view, we estimate that our PWV measurement precision is $\pm0.11$ mm. We explore the statistics of PWV variations over a range of timescales from less than an hour to days. We find that the amplitude of PWV variations within an hour is less than 1 mm for most (96.5\%) APOGEE field visits. By considering APOGEE observations that are close in time but separated by large distances on the sky, we find that PWV is homogeneous across the sky at a given epoch, with 90\% of measurements taken up to 70$^{\circ}$ apart within 1.5 hr having $\Delta\,\rm{PWV}<1.0$ mm. Our results can be used to help simulate the impact of water vapor absorption on upcoming surveys at continental observing sites like APO, and also to help plan for simultaneous water vapor metrology that may be carried out in support of upcoming photometric and spectroscopic surveys. 
\end{abstract}


\section{Introduction}

Ground-based astronomical observations at optical and near-infrared wavelengths suffer from wavelength-dependent absorption by molecules in Earth's atmosphere. These ``telluric'' absorption features, primarily due to H$_2$O, CO$_2$, O$_2$, and CH$_4$, may be particularly problematic for measurements where high precision is required, such as exoplanet transit photometry and stellar radial velocity measurements. Prominent bands of H$_2$O and O$_2$ at wavelengths less than 1 $\mu$m render portions of the optical spectrum virtually unusable due to the large optical depths of these lines, while the micro-telluric lines of H$_2$O, with depths less than 1$\%$, are problematic because they are so numerous. Absorption features due to water vapor can systematically bias radial velocity measurements \citep{cunha2014, fischer2016} and impose fundamental limits on differential photometry of cool stars \citep{blake2008,berta2012,baker2017}. For the case of H$_2$O, these problems are compounded by the fact that the quantity of water vapor above an observatory changes with time. The quantity of water vapor is usually reported as Precipitable Water Vapor (PWV), expressed in mm, which is the amount of rain that would result from all of the water vapor precipitating out of the atmosphere. 

There are several ways to measure PWV in real time. Narrow-band stellar photometry targeting specific wavelength regions containing water vapor lines can be used to determine the depths of the underlying telluric features, and therefore measures PWV directly. This type of approach is effective at measuring PWV with an accuracy of a few percent \citep{brooks2007,stubbs2007, ting2013, baker2017}. In the 1990s, it was shown that signal timing information from a Global Positioning System (GPS) receiver can be used to measure PWV (e.g. \citealt{bevis1992}). Water vapor concentration affects the index of refraction of the atmosphere at GPS radio frequencies (e.g., 1.2 and 1.6 GHz), and therefore is a major source of variations in the signal travel time between satellites and a receiver on Earth's surface. This technique is widely used in the atmospheric sciences community\footnote{\url{http://www.suominet.ucar.edu/}.} to study the behavior of water vapor in Earth's atmosphere, and an astronomical application of GPS-based PWV measurements is described in \cite{blake2011}. The water vapor content of the atmosphere can also be actively probed using a microwave radiometer, as in \cite{querel2014}. Finally, the depths of telluric water features imprinted on astronomical spectra can be modeled to directly assess the PWV at the time of a given observation (e.g. \citealt{querel2011}). In \cite{blake2011} this technique was directly calibrated against contemporaneous GPS-based PWV measurements at Apache Point Observatory (APO), home to the Sloan Digital Sky Survey (SDSS). 

Upcoming surveys like the Large Synoptic Survey Telescope \citep{lsst2008} and spectroscopic instruments like NEID \citep{neid2016} will require PWV information in order to extract the most information from their observations. While it is clear that PWV can change dramatically between nights and throughout the year, the short-term variability, as well as variability across the sky at a given time, are both less well characterized. \cite{querel2014} found that PWV is very homogeneous across the sky at Paranal in Chile, an exceptionally dry site \citep[median $\sim$2.4 mm;][]{kerber2015}. This implies that a single measurement of PWV to zenith could provide sufficient information to help mitigate the impact of water vapor absorption on many types of astronomical measurements. 

We investigate the temporal and spatial variability of water vapor at a lower altitude site with conditions representative of those found at many observatories around the world. We use about 400,000 high-resolution, near-infrared ($H$-band) spectroscopic observations of telluric standard stars from the Apache Point Observatory Galactic Evolution Experiment (APOGEE; \citealt{majewski2017}) to quantify the temporal and spatial variability of PWV at APO (32\degr~46\arcmin~49\arcmin~N, 105\degr~49\arcmin~13\arcmin~W, 2788 m a.s.l.). APOGEE is a multi-object spectrometer that gathers spectra of up to 300 objects over an area of 3$^\circ$ in diameter simultaneously. The APOGEE observing strategy results in a data set that probes temporal variability of PWV on timescales as short as $\sim$8 minutes (500 s) and spatial variability across both the 3$^\circ$ diameter of an APOGEE field and across wider regions of the sky. We find that the median PWV is 3.2 mm at APO. The large, short timescale PWV variations are uncommon and the typical variation in PWV at APO is less than 0.1 mm hr$^{-1}$. We find no evidence for strong PWV variations over degree scales and also find that measured PWV depends only weakly on altitude and azimuth for observations taken close in time. In Section 2 we describe the APOGEE data and our fits to the telluric standard spectra. In Section 3 we describe the statistical properties of our PWV measurements. In Sections 4 we summarize the main conclusions from this work and discuss how these results might benefit upcoming surveys.

\section{Data Collection and Reduction}
\subsection{Telluric spectra in APOGEE DR 13}

The APOGEE instrument on the Sloan Digital Sky Survey (SDSS) telescope at APO is capable of obtaining high-resolution ($R\approx22,500$) spectra of up to 300 objects simultaneously with fibers plugged into a plate covering an area approximately 3$^\circ$ in diameter. The spectra cover the $H$-band (1.52--1.7 $\mu$m) using a mosaic of three infrared detectors. The primary science goals of the APOGEE survey include large-scale studies of galactic kinematics and chemical composition. At infrared wavelengths, APOGEE is able to obtain high signal-to-noise ratio (SNR) spectra of giant stars in highly extincted regions of the galactic disk to large distances. However, this wavelength range suffers from substantial telluric absorption due to water vapor, CO$_2$, and CH$_4$, and therefore telluric standard stars were observed during every APOGEE observation to allow correction for these features.

We analyzed data from APOGEE Data Release 13 (DR13 - \citealt{sdssdr13}), which includes 2349 ``visits'' of 436 distinct fields between 2011 and 2014. In a typical visit of a given field, 35 fibers were placed on telluric standards, mostly hot A stars, spread across the 3$^\circ$ field of view of each plate and spatially distributed as evenly as possible. The typical integration time of a visit is 4000 s (i.e., with eight 500 s exposures). On each observing night, up to eight fields were visited, yielding thousands of spectra tracing spatial and temporal variations of telluric spectral features.

The reduction of the APOGEE data is described in \cite{nidever2015}. Raw data from each 500 s exposure were reduced by the APOGEE Visit Data Reduction (APRED) pipeline\footnote{\url{http://www.sdss.org/dr13/irspec/apred/}}, which generated 300 1D spectra (i.e., one spectrum per fiber). These 1D spectra are dark-subtracted, flat-fielded, and matched with a wavelength solution, flux errors, and a bad-pixel mask. No further correction was done at this stage, and thus the 1D spectra (referred to as ``ap1D'' spectra hereafter) contain all sky emission and absorption features. In the next step, all 500 s exposures from one visit were stacked to form a final 1D spectrum for each fiber. Sky and telluric features were modeled and removed, and flux calibration was performed in this step. The best-fit telluric model (i.e., a normalized spectrum of the telluric features; referred to as an ``apVisit'' telluric spectrum hereafter) derived by the APRED pipeline was attached to each apVisit spectrum.

Both ap1D and apVisit data are available through the SDSS Science Archive Server\footnote{\url{https://data.sdss.org/sas/dr13/apogee/}} (SAS). We identified telluric standard observations in the raw data using the $\rm{APOGEE\_TARGET2}$ field in the APOGEE targeting bitmask\footnote{\url{http://www.sdss.org/dr13/irspec/targets/}}. We collected all telluric spectra delivered during the three-year science operations of APOGEE, including 78,524 apVisit telluric spectra and 667,386 ap1D spectra. Using the mean (uncalibrated) flux of 1000 ADU per pixel (which roughly corresponds to a SNR of 50) in the ap1D spectra, we selected a subset of ``good'' data that include 44,240 apVisit and 353,920 ap1D spectra from 2,243 visits over 422 nights.

\subsection{Spectral fitting}
We focus our analysis on $\rm H_{2}O$ bands between 1.5146 $\micron$ and 1.5240 $\micron$ in the APOGEE spectral window. This wavelength range was chosen to contain only telluric water vapor features, avoiding low-level CO$_2$ and CH$_4$ features. To estimate the amount of water vapor in the atmosphere, we fit each (normalized) ap1D or apVisit telluric spectrum to a telluric spectral template multiplied by a scale factor (SF) to account for variations in the water vapor optical depth
\begin{equation}
\rm ln \it F(\lambda) = \rm SF\cdot ln [\it T(\lambda)\otimes\rm LSF].
\label{eq:fit}
\end{equation}
Here, $F(\lambda)$ is the scaled template to be compared with the data, $T(\lambda)$ is the original template, LSF is the line spread function, and $\otimes$ indicates convolution. We used the TAPAS web interface\footnote{\url{http://ether.ipsl.jussieu.fr/tapas/}} to generate telluric absorption models appropriate for APO using an average latitude winter atmospheric model \citep{tapas2014}. The telluric template contains only water vapor. The resolution of the original TAPAS template is $10^{-6}$ $\micron$, so we resampled the template (after convolution) onto the APOGEE wavelength grid before calculating the SF.

Note that the APOGEE LSF is wavelength- and fiber-dependent \citep{nidever2015}. Using a few prominent sky emission lines within the chosen range of wavelength, we find that the LSF is extremely stable during most visits. The relative change in the line profile (as measured by the FWHM) among eight consecutive 500 s exposures is of the order of $10^{-2}$, thus having negligible influence on the derived SF compared with other sources of uncertainties in the fit. For this reason, we applied the same LSF model, which is the one provided for each apVisit spectrum, for all ap1D spectra within a visit.

All ap1D spectra must be properly normalized to have a flat stellar continuum before calculating the SF using Equation \ref{eq:fit}. Finding the continuum is an iterative process based on median-filtering. Strong sky emission lines, telluric lines, and hydrogen lines of the star are removed in the first few iterations. Then, asymmetric sigma-clipping is applied to make the fitted continuum trace the upper envelope of remaining pixels. We found that the continuum determined in this way (rather than symmetric sigma-clipping) makes our SFs derived from ap1D spectra consistent best with those computed for apVisit telluric spectra, which have already been normalized in the APRED pipeline.

An initial estimate of the SF for each ap1D or apVisit telluric spectrum was obtained by matching the total absorption by telluric lines in the observed spectrum (essentially the total equivalent width) to that of a scaled TAPAS template. Then, a brute force search for the best-fit SF was performed by calculating $\chi^2$ ($=\sum(\left[DATA-F(\lambda)\right]/\sigma)^{2})$ over a very fine grid around the initial SF estimate. A parabola was fitted to the resulting $\chi^2$ curve to determine the SF that yielded the minimum $\chi^2$. In the search for the best-fit SF, only pixels within 5 pixels of a telluric line with normalized depth greater than 0.01 in the convolved telluric template (i.e., $F(\lambda)$ in Equation \ref{eq:fit}) were used for calculating $\chi^2$. Bad pixels flagged in $\rm{APOGEE\_PIXMASK}$ were also masked out\footnote{\url{http://www.sdss.org/dr13/algorithms/bitmasks/}}. Even under perfectly stable and homogeneous conditions, the water vapor optical depth is a function of path length through Earth's atmosphere, $\rm sec\it(z)$, where $z$ is the zenith angle. To account for this effect, the best-fit SF for each spectrum is divided by airmass to create a quantity we refer to as the Reduced Scale Factor (RSF). 

\subsection{Calibrating RSF with GPS data}
It has been demonstrated that a multi-band, geodetic-quality GPS receiver and a high-accuracy barometer can be combined to derive PWV \citep{bevis1992}. In North America, a network of GPS-based PWV monitors is maintained by UCAR (University Corporation for Atmospheric Research), with a station operated by UNAVCO located at APO. The PWV monitor at APO is equipped with a Trimble\textsuperscript{\textregistered} NetRS\textsuperscript{\textregistered} GPS receiver with a Trimble\textsuperscript{\textregistered} Choke Ring Antenna, while a Vaisala WXT510 weather transmitter is used to monitor the ambient temperature and pressure.

\begin{figure}
\begin{center}
\includegraphics[width=1.0\textwidth]{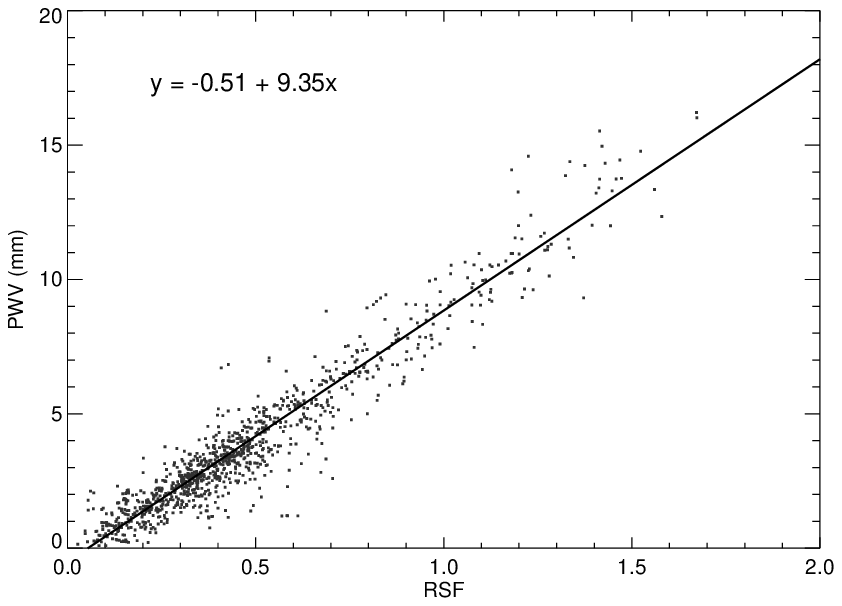} 
\caption{Linear correlation between PWV measured by the GPS monitor at APO and the RSF derived from APOGEE spectra. Each plotted RSF is the mean value for all 35 apVisit telluric spectra of telluric standard stars within one plate, whereas the PWV is the mean value of all measurements reported by the GPS monitor during the observation. The solid line is the best least-square fit to the data. The Pearson correlation coefficient is 0.96. The RMS scatter about the best-fit line is 0.84 mm and is dominated by uncertainty in the GPS-based measurements. 
}
\label{fig:PWV_vs_RSF}
\end{center}
\end{figure}

Using SuomiNet\footnote{\url{http://www.suominet.ucar.edu. APO site ID: P027}}, we downloaded historical data recorded at the APO station. Except for a gap between 2013 January 1 and 2013 November 14, PWV was measured regularly at 30-minute intervals between 2011 and 2014. Figure \ref{fig:PWV_vs_RSF} shows that the RSF we derived is linearly correlated with the GPS data (Pearson correlation coefficient = 0.96, residual RMS = 0.84) 

\begin{equation}
\rm PWV~(mm) = 9.35\cdot\rm RSF-0.51.
\label{eq:gps}
\end{equation}
The accuracy of GPS-based zenith PWV measurements has been assessed by comparison to contemporaneous radiometer measurements. The uncertainty is typically 1 mm and is dominated by uncertainty in the temperature profile of Earth's atmosphere \citep{gpserror}. It appears that this level of uncertainty is consistent with the scatter observed in Figure \ref{fig:PWV_vs_RSF}. 

\subsection{Uncertainties and biases}
\begin{figure}
\begin{center}
\includegraphics[width=1.0\textwidth]{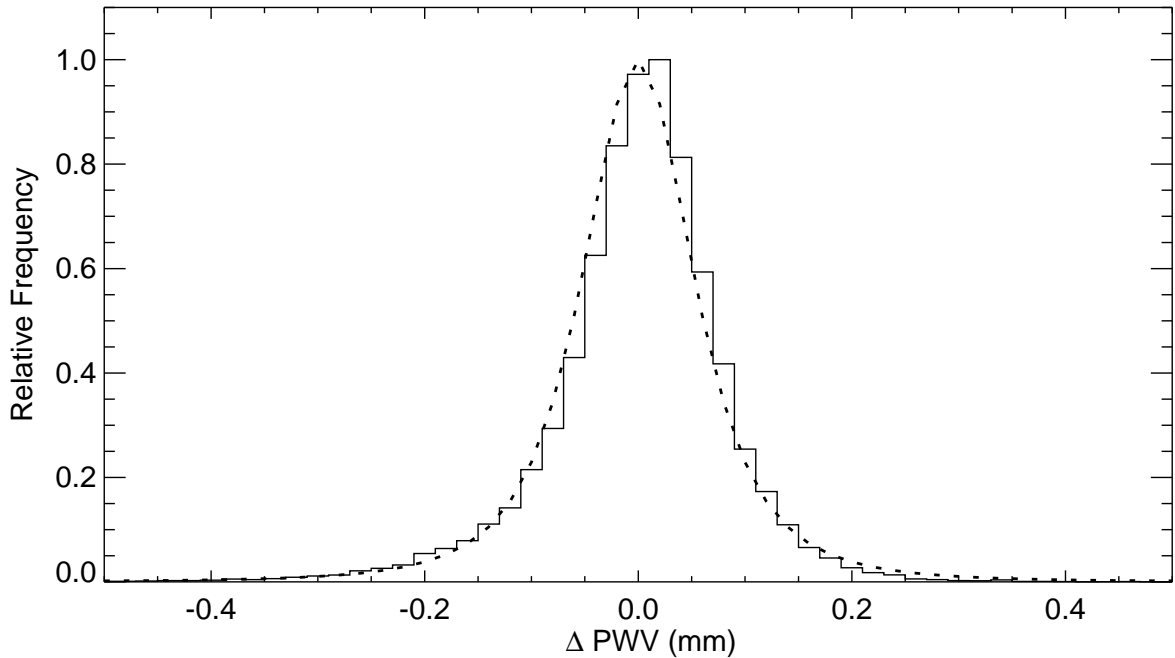} 
\caption{Empirical error distribution for PWV estimates based on fits to apVisit telluric spectra. For any given spectrum, $\Delta$ PWV is the difference between PWV derived from that spectrum and the plate average PWV for the visit. The dashed line is the best-fit Moffat function approximation of this distribution ($\chi^2_{\nu}=22.9$, $\nu=19$). Note that the best-fit function is forced to be centered on $\Delta$ PWV = 0.0 mm. The standard deviation of the Gaussian core of the distribution is 0.06 mm.}
\label{fig:apVisit_error_dist}
\end{center}
\end{figure}

Assuming that PWV is always uniform across the field of view of APOGEE, the relative scatter of PWV measured with different fibers can be used as an empirical estimate of the statistical errors on the SF fits. Distributions of the relative scatter of PWV for ap1D and apVisit telluric spectra are shown in Figures \ref{fig:ap1D_error_dist} and \ref{fig:apVisit_error_dist}. 32,961 apVisit and 257,864 ap1D spectra from 1,236 visits on 417 nights were used to generate the two plots (only visits that have at least 10 fibers, for which all ap1D spectra have SNRs greater than 50, are considered).

In both cases, we find that the empirical SF error distribution has larger tails than a Gaussian distribution, but is well fit by a Moffat distribution. For ap1D data, the best-fitting Moffat profile yields a $\chi^2_{\nu}$ of 11.7 with degrees of freedom ($\nu$) of 19. For apVisit data, $\chi^2_{\nu}=22.9$ and $\nu=19$. Considering only the Gaussian core of these distributions, we find that the statistical error in our SF fits is $\pm0.11$ mm for PWV derived from the APOGEE ap1D spectra, or $\pm0.06$ mm for apVisit telluric spectra, both being substantially less than the GPS-based PWV uncertainty ($\sim$1 mm RMS; \citealt{gpserror}). Because the SNR of an apVisit telluric spectrum is generally 3 to 4 times higher than that of the corresponding ap1D spectra, the lower SF fit error for apVisit data is as expected. 

We also carried out Monte Carlo simulations to estimate the impact of flux errors on the SF uncertainties. We began with a simulated spectrum stacked from 2000 high-SNR (SNR$>$400) ap1D spectra. Gaussian noise (with a varying amplitude to simulate a varying SNR) was added to the spectrum, which was then used as the input for our fitting procedures to derive the SF. Under each SNR setting, we run the simulation 200 times. We found that the statistical uncertainty in PWV measured with the simulated ap1D spectra is $\pm$0.1 mm for SNR = 100 (i.e., approximately the median SNR of the data set), which is in good agreement with the empirical estimate.

\begin{figure}
\begin{center}
\includegraphics[width=1.0\textwidth]{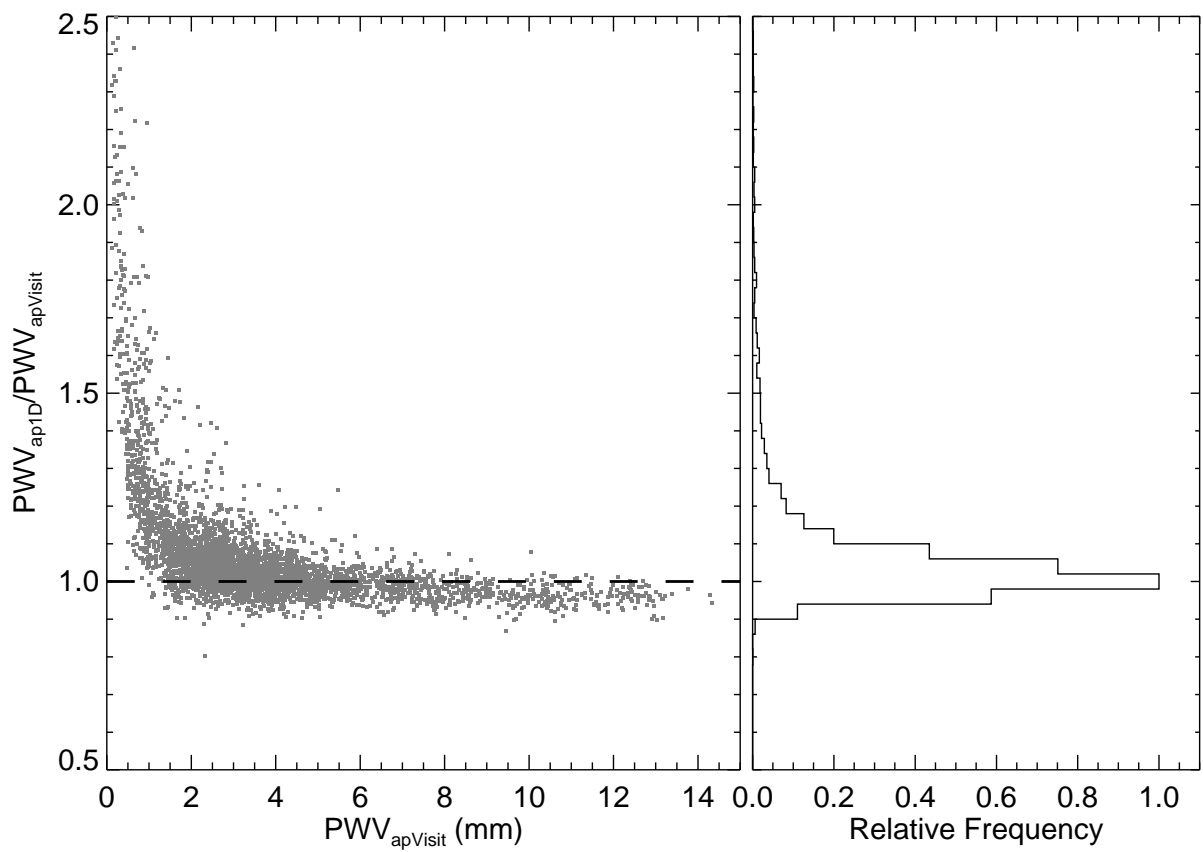} 
\caption{Ratio of mean PWV derived from ap1D spectra comprising a visit to that derived from the apVisit telluric spectrum produced by the APOGEE data pipeline. Each point represents a visit of a given telluric standard star. The peak near unity indicates that the PWV derived from the ap1D spectra are unbiased relative to those derived from apVisit telluric spectra.
}
\label{fig:ap1D_vs_apVisit}
\end{center}
\end{figure}

We find consistent results when comparing PWV estimates from the apVisit telluric spectra to those derived from the ap1D spectra from which the apVisit spectra are formed. Plotted in the left panel of Figure \ref{fig:ap1D_vs_apVisit} is the ratio of the mean PWV derived from eight ap1D spectra of a given fiber and visit to the PWV derived from the corresponding apVisit telluric spectrum. The distribution of this ratio narrowly peaks at unity, with 80\% measurements falling between 0.9 and 1.1. There are, however, some biases noticeable at very low (i.e., PWV $<$ 1 mm) or very high (i.e., PWV $>$ 10 mm) PWV values where our fits to the ap1D spectra may overestimate or underestimate the PWV relative to the apVisit values. A closer inspection of individual ap1D spectra taken at very low or very high PWV suggests that these biases may be introduced by our continuum fitting process. These instances only represent a small portion (10.4\%) of the entire data set.  Nevertheless, since ap1D spectra from the same visit and the same fiber are similar to each other (in terms of their fluxes or SNRs), even though their mean PWV might be over- or under-estimated, relative changes in the PWV between individual exposures within a visit should still be accurate and reliable. Because we will primarily use apVisit data to probe telluric variations between visits or nights, whereas ap1D data will mainly be used to study short-timescale (i.e., within a visit) variations, the small biases corresponding to unusual observing conditions seen in Figure \ref{fig:ap1D_vs_apVisit} affect our subsequent analyses negligibly.

We note that the APRED pipeline also derived scale factors for ap1D spectra, and their values (and reduced ap1D spectra) can be found in APOGEE apCframe files. However, we chose to re-fit all 1D spectra instead of using the APRED results because it allowed us to examine the fitting process and performance (e.g., the goodness and robustness of the fit) more carefully. It also makes it more convenient to expand our study to include different telluric models and/or more molecules in the future. We cross-checked the APRED scale factors and ours and found that they are consistent with each other.

\section{Analysis}
\subsection{Long-term PWV Variations}
\begin{figure}
\begin{center}
\includegraphics[width=1.0\textwidth]{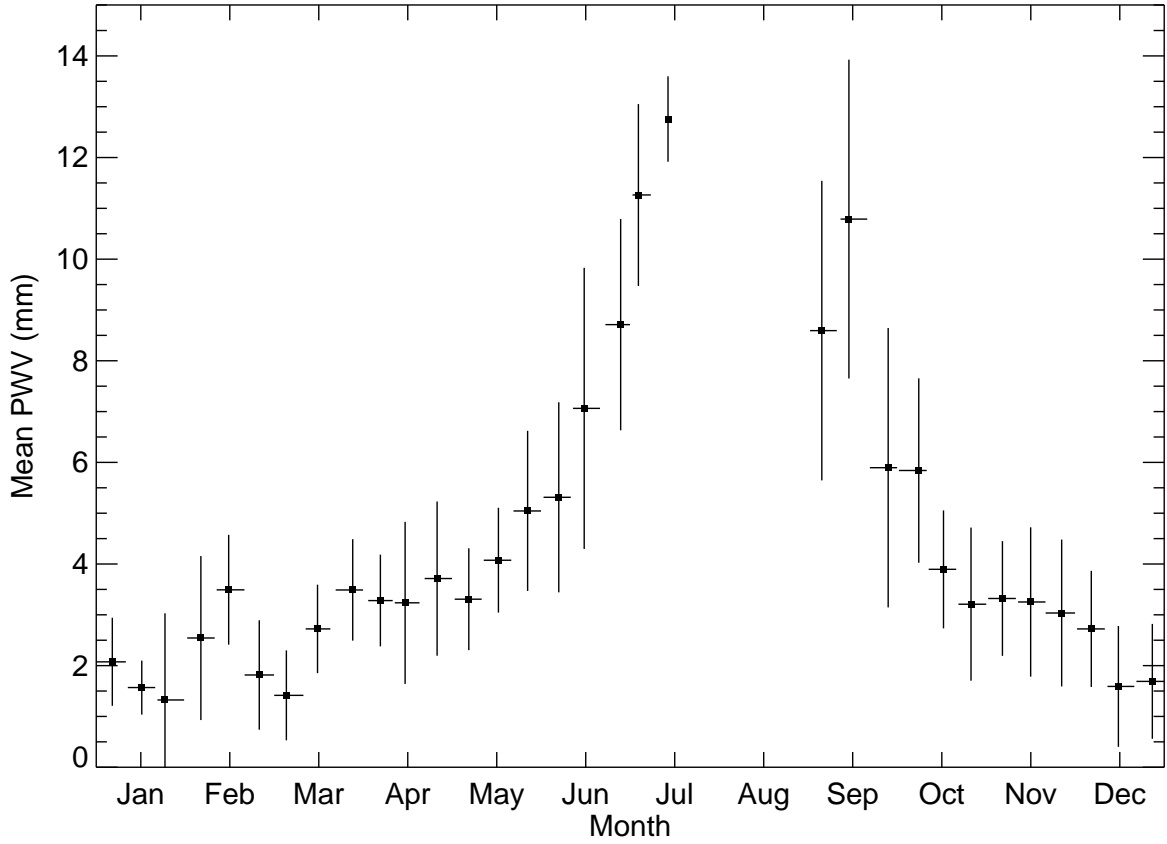} 
\caption{Mean PWV of each 10-day period. Vertical error bars are standard deviations in PWV. Based on 44,240 apVisit telluric spectra taken on 422 nights over 32 months between 2011 September and 2014 July (no observation in late-July and August).
}
\label{fig:pwv_seasonal}
\end{center}
\end{figure}

We can investigate the long-term behavior of PWV at APO by using per-visit estimates that correspond to the mean PWV derived from all telluric spectra in a single epoch, which typically lasted for $\sim$4,000 s. Seasonal variations in PWV are clearly seen in our data (Figure \ref{fig:pwv_seasonal}). The average PWV observed between June and September is a few times higher than that in the winter. PWV shows a steep rise between May and June, followed by a rapid decrease between September and October. 

\begin{figure}
\begin{center}
\includegraphics[width=1.0\textwidth]{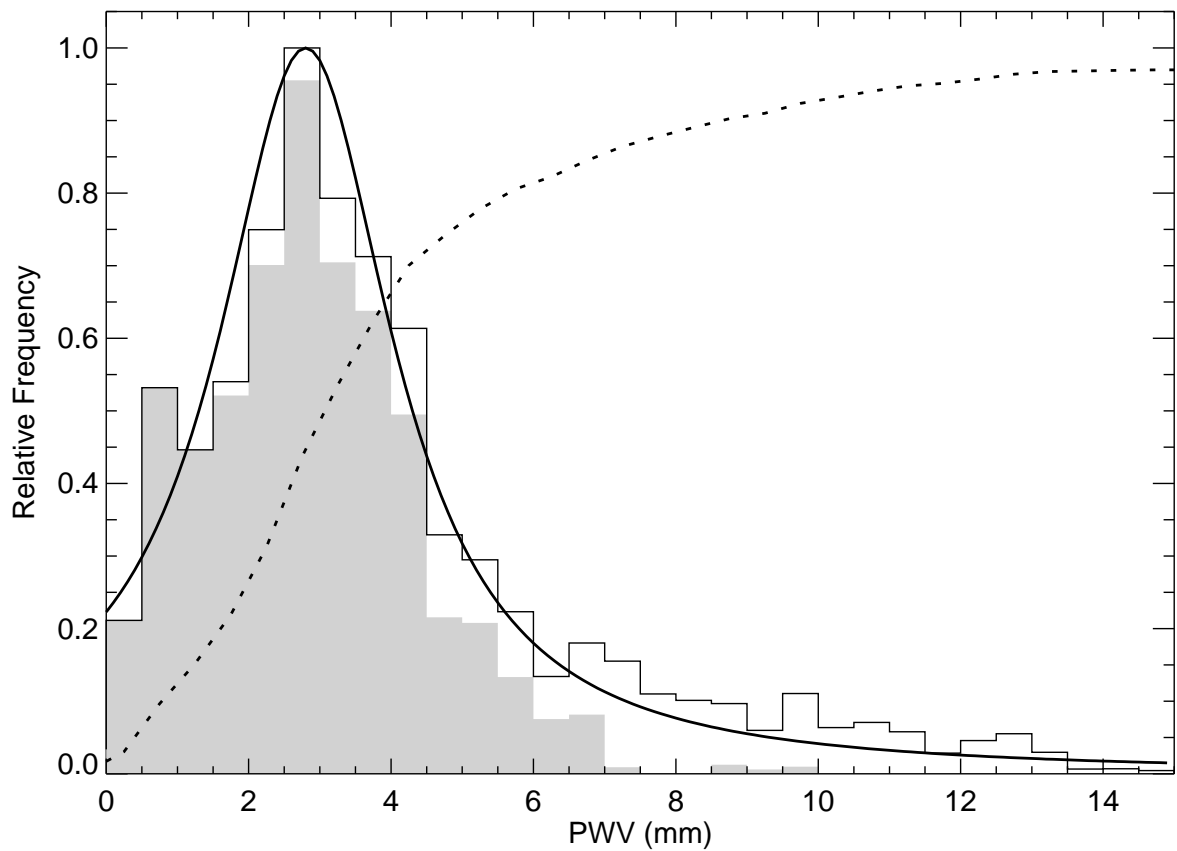} 
\caption{Distribution of mean PWV per visit. Based on the same set of apVisit spectra used in Figure \ref{fig:pwv_seasonal}. The dotted line is the cumulative distribution, and the solid line is a truncated Lorentz profile that we found to be a good match to the observed distribution ($\chi^2_{\nu}=59.4$, $\nu=27$). The histogram in grayscale shows the PWV distribution for a subset of data without observations between June and September.
}
\label{fig:pwv_histogram}
\end{center}
\end{figure}

We find that the distribution of PWV (Figure \ref{fig:pwv_histogram}) has a long tail, which is almost entirely due to observations conducted between June and September. Empirically, we find that the distribution of PWV is well matched by a (truncated) Lorentz distribution ($\chi^2_{\nu}=59.4$, $\nu=27$) with the location parameter 2.8 mm and the scale parameter 1.5. While we do not assign physical significance to this parameterization of the PWV distribution, we hope that it might be useful for future efforts to simulate the impact of PWV on astronomical surveys. Taking into account all data, the mean (median) PWV at APO is 3.8 (3.2) mm, with PWV lower than 10 mm for 92\% of epochs. Excluding data taken between June and September, the mean (median) PWV reduces to 2.9 (2.8) mm. 

\subsection{Short-term PWV Variations}
By considering the individual ap1D spectra taken over the course of a single visit, as well as visits to different fields throughout a single night, the APOGEE data enable us to investigate the variation of PWV over a range of timescales from minutes to hours. 

\begin{figure}
\begin{center}
\includegraphics[width=1.0\textwidth]{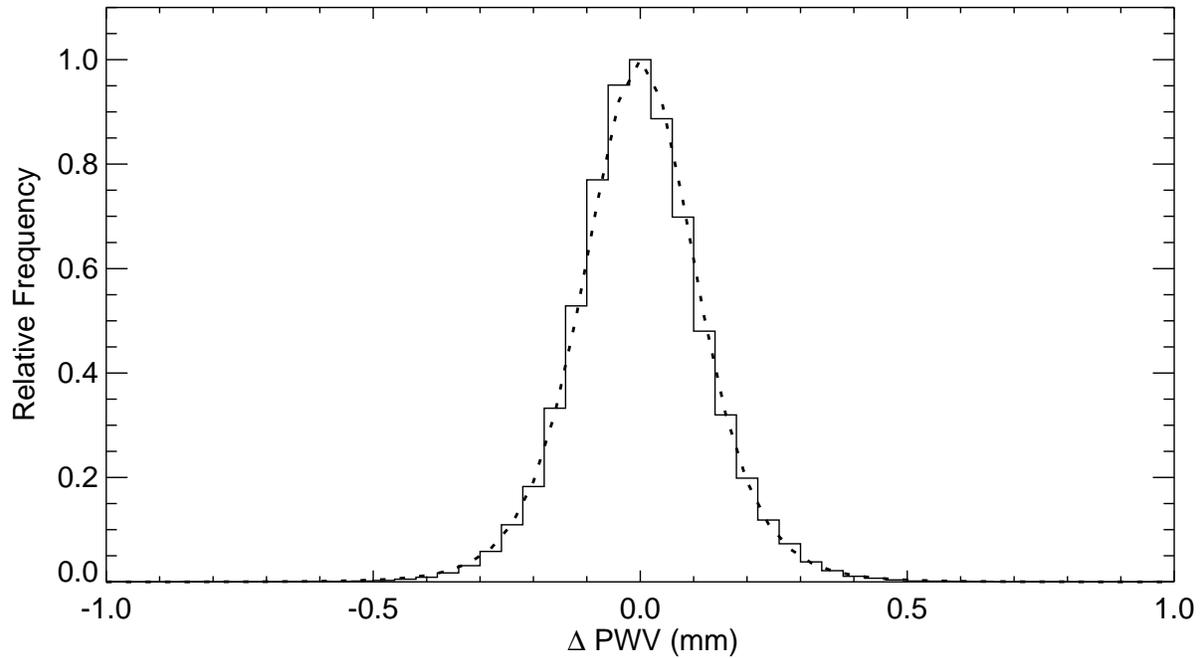} 
\caption{Empirical error distribution for PWV estimates based on fits to ap1D spectra. For any given spectrum, $\Delta$ PWV is the difference between PWV derived from that spectrum and the plate average PWV for that single exposure. The dashed line is the best-fit Moffat function approximation of this distribution ($\chi^2_{\nu}=11.7$, $\nu=19$). Note that the best-fit function is forced to be centered on $\Delta$ PWV = 0.0 mm. The standard deviation of the Gaussian core of the distribution is 0.11 mm.}
\label{fig:ap1D_error_dist}
\end{center}
\end{figure}

\begin{figure}
\begin{center}
\includegraphics[width=1.0\textwidth]{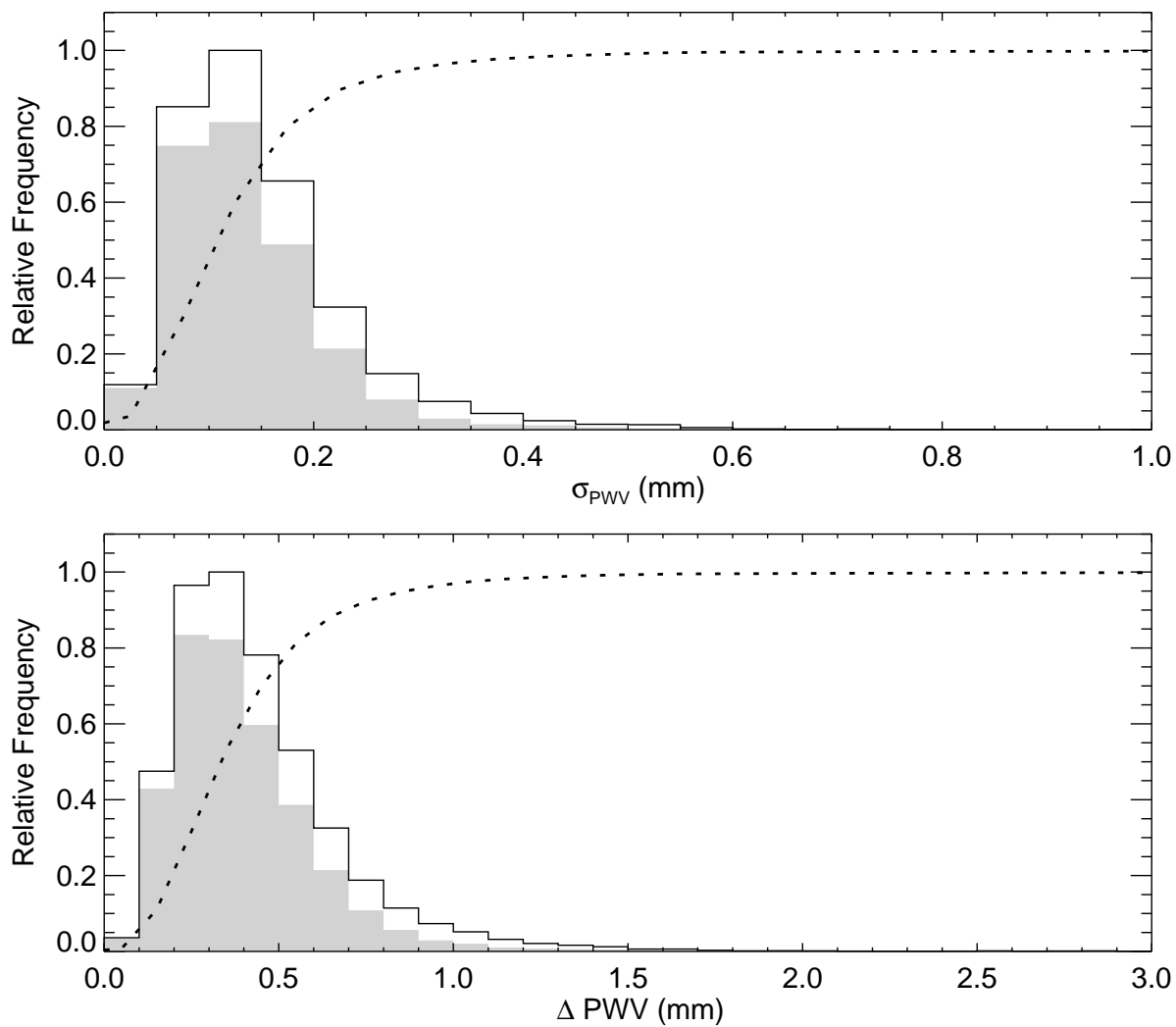} 
\caption{Variation of PWV within a visit. The distributions are based on ap1D spectra, and thus contain any temporal variations in PWV on timescales up to $\sim$1 hr and spatial scales up to 3$^{\circ}$. 353,920 ap1D spectra from 422 nights are used in this figure. Upper panel: the standard deviation of PWV measured with eight consecutive 500 s exposures through a given fiber. Lower panel: the peak-to-valley spread of PWV measured with eight consecutive 500 s exposures through a given fiber. Dotted lines are cumulative distributions. Grayscale histograms show PWV variations for a subset of data that exclude spectra taken between June and September.
}
\label{fig:ap1d_variation}
\end{center}
\end{figure}

The PWV variation (quantified by the standard deviation $\sigma_{\rm PWV}$ and the peak-to-valley spread $\Delta$ PWV) derived from consecutive ap1D spectra of the same star in a given visit, which is typically 4000s in total duration, is plotted in Figure \ref{fig:ap1d_variation}. The distribution of $\sigma_{\rm PWV}$ peaks around 0.1--0.15 mm, consistent with the empirical PWV fit uncertainty shown in Figure \ref{fig:ap1D_error_dist}, though with a long tail. However, our analysis shows that the distribution of $\Delta$ PWV has a longer tail than expected from the empirical error distribution shown in Figure \ref{fig:ap1D_error_dist}, indicating that low-level PWV variations on hour timescales are present, though typically with amplitudes less than our measurement precision. We attempted to simulate the distribution of $\Delta$ PWV shown in the bottom panel of Figure \ref{fig:ap1d_variation} as the convolution of the empirical ap1D PWV error distribution and a small linear trend in PWV during each visit with slope drawn from a Gaussian distribution. We found that the best fit PWV slope distribution was $N$[0, 0.06], where the sigma of the Gaussian is the slope of the PWV change in units of mm 500 s$^{-1}$. We conclude that the typical variation in PWV at APO over the course of an hour is small compared to our measurement precision, less than 0.1 mm hr$^{-1}$. However, the tail of the distribution shown in bottom panel of Figure \ref{fig:ap1d_variation} indicates that $\Delta$ PWV $>$ 1 mm in an hour can occur (in approximately 3.5\% of all visits). More statistics of $\sigma_{\rm PWV}$ and $\Delta$ PWV based on ap1D data are presented in Table \ref{tab:ap1d_statistics}.

\begin{figure}
\begin{center}
\includegraphics[width=1.0\textwidth]{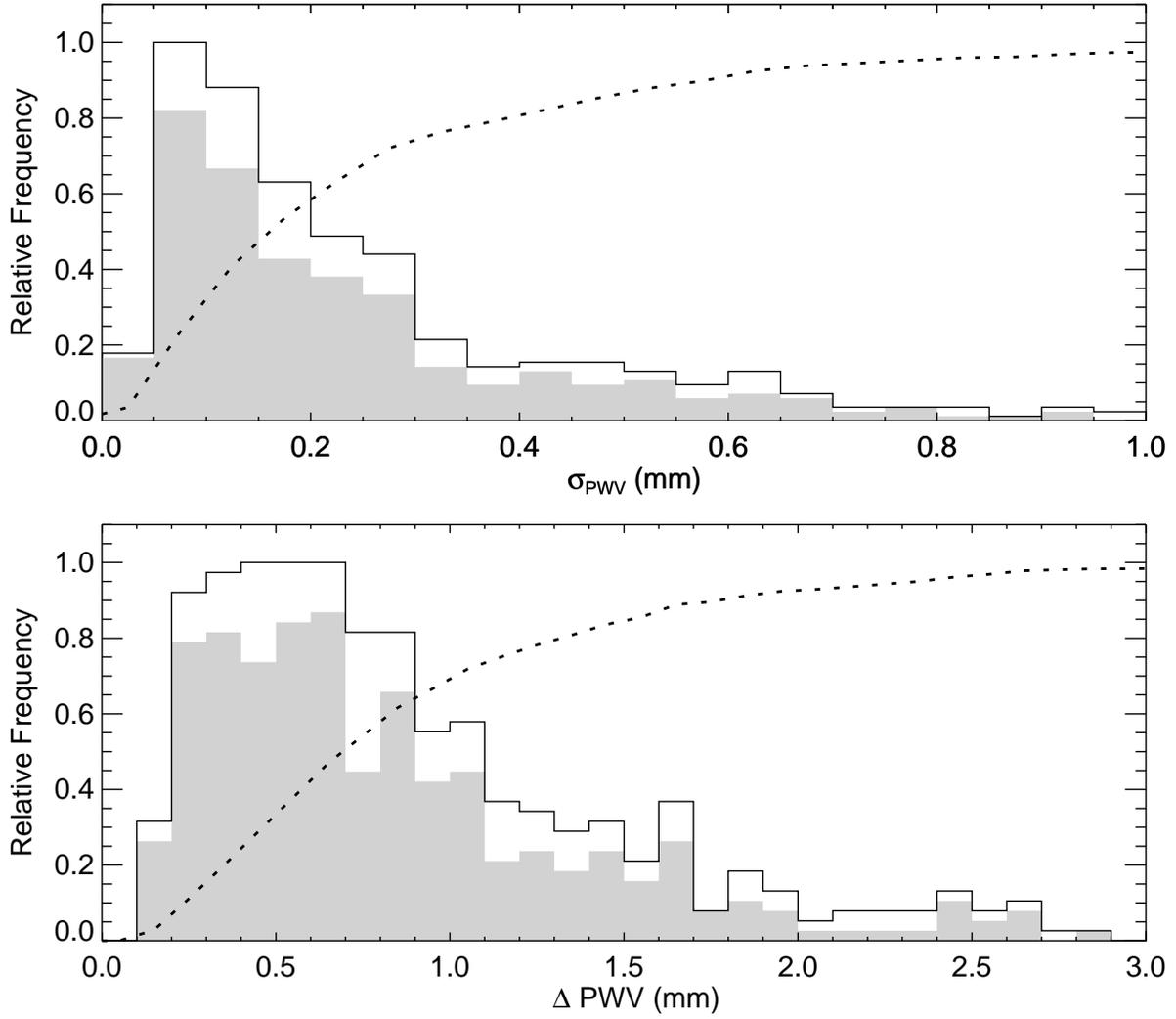} 
\caption{Variations of PWV within a night. The distributions are based on apVisit telluric spectra, and thus contain any temporal variations in PWV on timescales from an hour to $\sim$8 hr and spatial scales up to 70$^{\circ}$. 44,240 apVisit telluric spectra from 422 nights are used in this figure. Upper panel: the standard deviation of PWV measured with all apVisit telluric spectra for a given night. Lower panel: the peak-to-valley spread of PWV measured with all apVisit telluric spectra for a given night. Dotted lines are cumulative distributions. Grayscale histograms show PWV variations for a subset of data that exclude spectra taken between June and September.
}
\label{fig:apVisit_variation}
\end{center}
\end{figure}

Similarly, PWV variations on timescales from hours up to a night can be characterized by the apVisit data (Figure \ref{fig:apVisit_variation}). The distribution of $\sigma_{\rm PWV}$ and $\Delta$ PWV (of different epochs in a single night) both have long, slow-decreasing tails. We find that PWV is still generally quite stable on this timescale, with $\sigma_{\rm PWV}<0.3$ mm within 70\% of nights, or $\Delta$ PWV $<$ 1 mm within 65\% of nights. PWV variations greater than 2 mm occur during approximately 7\% of nights. More statistics of $\sigma_{\rm PWV}$ and $\Delta$ PWV based on apVisit data can be found in Table \ref{tab:apVisit_statistics}.

\begin{figure}
\begin{center}
\includegraphics[width=1.0\textwidth]{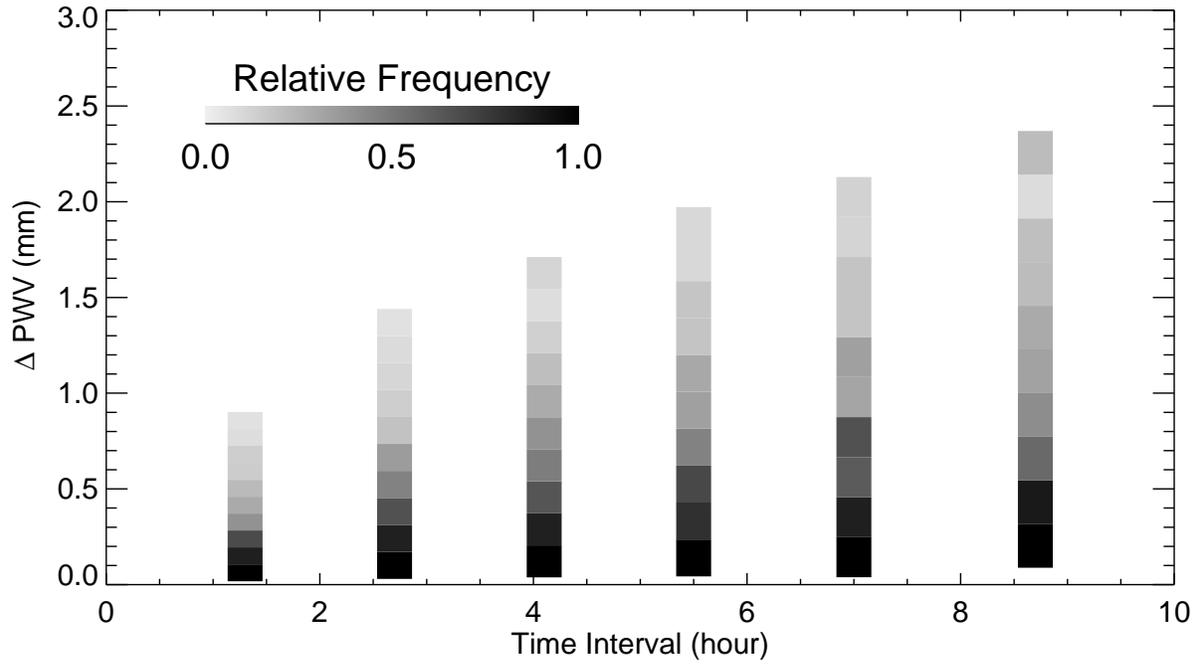} 
\caption{The 90-percent spread (peak to valley) of $\Delta$ PWV as a function of the time between two observations. Relative frequency within each bin is shown in grayscale. 4971 field-pairs are used in this plot.
}
\label{fig:temporal_variations}
\end{center}
\end{figure}

\begin{figure}
\begin{center}
\includegraphics[width=1.0\textwidth]{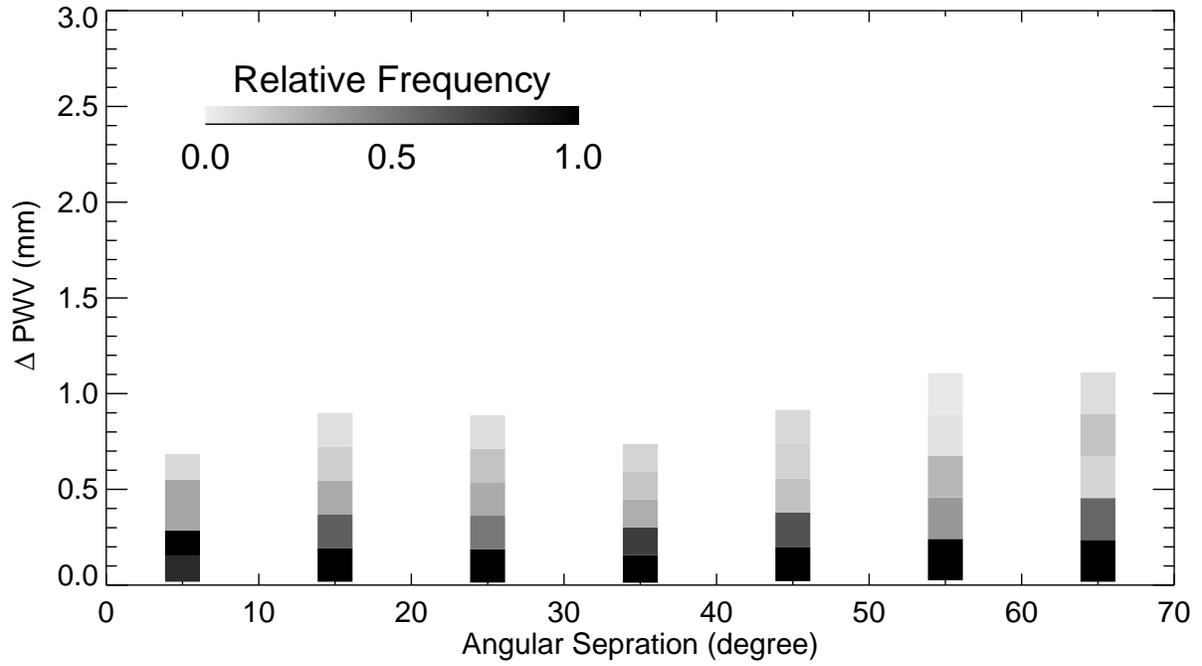} 
\caption{The 90-percent spread (peak to valley) of $\Delta$ PWV as a function of the angular separation on the sky between two consecutive observations taken within 1.5 hours of each other. Relative frequency within each bin is shown in grayscale. 1503 field-pairs are used in this plot.
}
\label{fig:spatial_variations}
\end{center}
\end{figure}

\begin{figure}
\begin{center}
\includegraphics[width=1.0\textwidth]{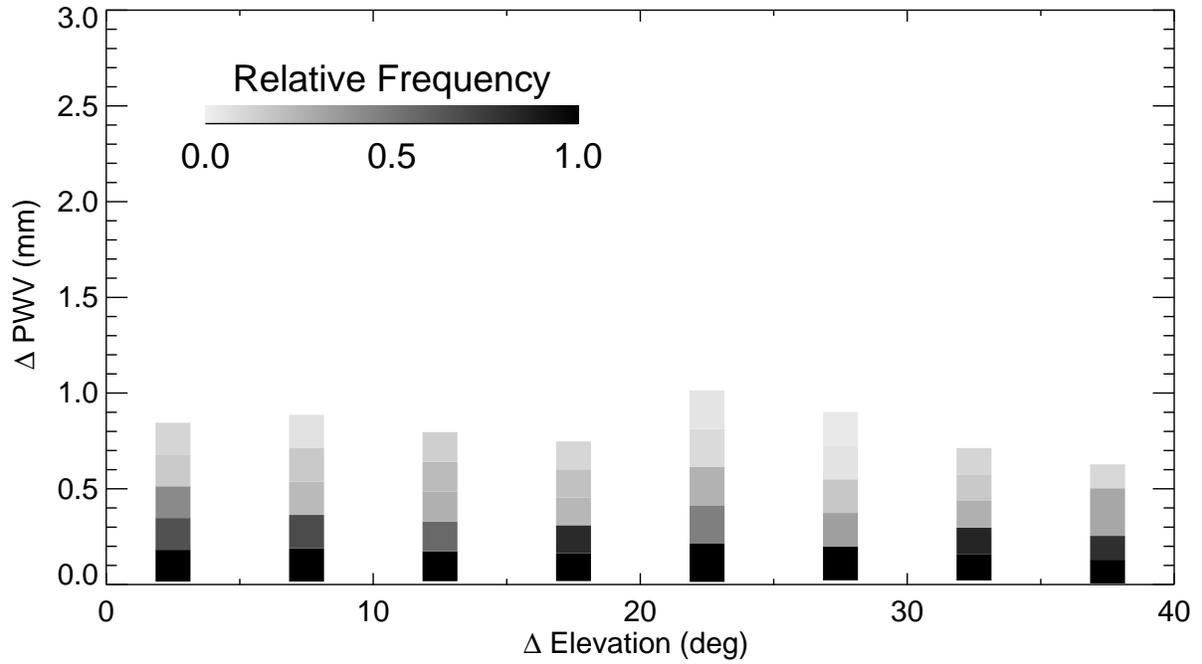} 
\caption{The 90-percent spread (peak to valley) of $\Delta$ PWV as a function of the change in elevation between two consecutive observations taken within 1.5 hours of each other. Relative frequency within each bin is shown in grayscale. 1,503 field-pairs are used in this plot. Because most APOGEE fields were observed between the elevation of 40$^\circ$ and 85$^\circ$, the change in elevation is generally less than 40$^\circ$. 
}
\label{fig:alt_variations}
\end{center}
\end{figure}

In a typical night, up to eight APOGEE fields were visited. Using these fields, which were temporally and spatially separated from each other, we can define a number of field-pairs, each of which tells us how PWV changes over a certain time interval and angular distance. In Figure \ref{fig:temporal_variations}, 90-percent spread (peak to valley) of PWV found between 4971 field-pairs are plotted. We find that large temporal variations in PWV are possible within a night, and that the likelihood of large relative PWV variations between observations increases with time. In Figure \ref{fig:spatial_variations}, we investigate the spatial variability by focusing on situations where APOGEE switched between two plates providing PWV measurements separated by up to 70$^\circ$ on the sky, but less than 1.5 hr in time (i.e., slightly longer than one typical visit). Using 1503 qualified field-pairs, we find that the spatial variation of PWV is very small, typically less than 0.5 mm (90-percent, peak to valley). Using the same set of field-pairs, we plot in Figure \ref{fig:alt_variations} the variability of PWV as a function of the change in elevation, and we found a similar result that the variation of PWV is very small, typically less than 0.5 mm, between consecutive observations at different elevation.

Using microwave radiometry, \citet{querel2014} reported spatial variation of PWV across the sky down to 27\fdg5 elevation with a median variation of 0.32 mm (peak to valley) or 0.07 mm (RMS) at Paranal, Chile. Our results suggest that at APO, a lower altitude site with conditions representative of those found at many observatories around the world, the water vapor behaves similarly.

\section{Discussion and Conclusions}
Using $\sim$400,000 high-resolution, near-infrared ($H$-band) spectroscopic observations of telluric standard stars from the APOGEE survey, we quantify the temporal and spatial variability of PWV at APO. We convert our measurements of the depths of water vapor absorption lines around 1.52~$\mu$m to PWV using simultaneous measurements from a GPS-based PWV monitor located at APO. Using simultaneous measurements over the 3$^\circ$ APOGEE FOV, we estimate that our typical statistical error on PWV is $\pm0.11$mm, significantly more precise than the GPS-based estimates (1 mm RMS). Given the nature of the APOGEE survey, we are able to probe PWV variation on a wide range of temporal and spatial timescales. We find that variations in PWV on timescales shorter than 1 hour are typically small, below our measurements precision. The GPS-based PWV monitoring station at APO provides measurements at 30-minute cadence, so our measurements probe a shorter-timescale regime at higher precision. We investigate the spatial dependence of PWV by considering APOGEE observations obtained close in time but separated by a large angle on the sky. We find that the angular dependence (as well as the elevation dependence) of PWV is very small, typically less than 0.5 mm (90-percent, peak to valley) even for sight lines separated by up to 70$^\circ$. Using microwave radiometry, \citet{querel2014} found that the spatial dependence of PWV across the sky at a given epoch is surprisingly small at Paranal in Chile, with peak-to-valley variations typically less than 0.5 mm. Our results suggest that at APO, a lower altitude site with conditions representative of those found at many observatories around the world, the water vapor is also spatially homogeneous.

Our results are useful for planning future efforts to monitor PWV as part of a wide range of astronomical surveys. For example, \cite{baker2017} calculated that high-precision differential photometry of cool stars for transiting exoplanet studies may require correcting for the impact of differential extinction by Earth's atmosphere. These authors found that if PWV is known well, to better than 1 mm, then photometric corrections can be calculated directly using atmospheric models. Upcoming extreme precision radial velocity surveys will benefit from contemporaneous measurements of water vapor optical depth as in input to models designed to minimize the impact of ``micro-telluric'' lines \citep{cunha2014}. Our analyses indicate that high-cadence PWV measurements made along a single sight line (toward Zenith, for example) are likely sufficient for a broad range of astronomical applications. 

\acknowledgments
We thank the anonymous referee for helpful comments that helped to improve this manuscript. This work was performed, in part, by S.H. under contract with the Jet Propulsion Laboratory (JPL) funded by NASA through the Sagan Fellowship Program executed by the NASA Exoplanet Science Institute. The GPS data used in the paper are based upon work supported by the National Science Foundation under Grant No. 0313588.

{\it Facilities:} \facility{Sloan}.

\bibliographystyle{apj}

\clearpage

\clearpage

\clearpage

\clearpage

\clearpage

\clearpage

\clearpage

\clearpage

\clearpage

\clearpage

\clearpage

\begin{deluxetable}{ccccccccc}
\tabletypesize{\scriptsize}
\tablecaption{Statistics of PWV variations per fiber per visit\label{tab:ap1d_statistics}}
\tablewidth{0pt}
\tablehead{
\colhead{Percentiles} & \colhead{0--1 mm} & \colhead{1--2 mm} & \colhead{2--3 mm} & \colhead{3--4 mm} & \colhead{4--6 mm} & \colhead{6--8 mm}  
& \colhead{8--10 mm} & \colhead{10--15 mm} \\
}
\startdata
10 & 0.05 & 0.06 & 0.06 & 0.07 & 0.08 & 0.09 & 0.10 & 0.12 \\
25 & 0.06 & 0.08 & 0.08 & 0.09 & 0.11 & 0.12 & 0.13 & 0.16 \\
\textbf{50} & \textbf{0.08} & \textbf{0.11} & \textbf{0.12} & \textbf{0.13} & \textbf{0.15} & \textbf{0.17} & \textbf{0.17} & \textbf{0.22} \\
75 & 0.12 & 0.15 & 0.16 & 0.18 & 0.20 & 0.25 & 0.23 & 0.30 \\
90 & 0.15 & 0.19 & 0.21 & 0.23 & 0.27 & 0.40 & 0.32 & 0.37 \\
 &  &  &  &  &  &  &  & \\
10 & 0.14 & 0.16 & 0.18 & 0.20 & 0.23 & 0.26 & 0.29 & 0.34 \\
25 & 0.18 & 0.23 & 0.25 & 0.27 & 0.31 & 0.35 & 0.39 & 0.45 \\
\textbf{50} & \textbf{0.26} & \textbf{0.32} & \textbf{0.35} & \textbf{0.38} & \textbf{0.43} & \textbf{0.50} & \textbf{0.50} & \textbf{0.63} \\
75 & 0.34 & 0.42 & 0.47 & 0.52 & 0.59 & 0.73 & 0.69 & 0.87 \\ 
90 & 0.42 & 0.54 & 0.62 & 0.68 & 0.77 & 1.11 & 0.91 & 1.11 \\
  &  &  &  &  &  &  &  & \\
Count & 1288 & 3147 & 4929 & 4742 & 4647 & 1799 & 1116 & 1065 
\enddata
\tablecomments{The standard deviation (Rows 1--5) and peak-to-valley spread (Rows 6--10) of PWV measured between consecutive exposures for a given fiber and visit, binned by the mean PWV of the visit. Based on the same set of data shown in Figure \ref{fig:ap1d_variation}. The median values are shown in boldface. All PWV values in millimeter.}
\end{deluxetable}

\clearpage

\begin{deluxetable}{ccccccccc}
\tabletypesize{\scriptsize}
\tablecaption{Statistics of PWV variations per night\label{tab:apVisit_statistics}}
\tablewidth{0pt}
\tablehead{
\colhead{Percentiles} & \colhead{0--1 mm} & \colhead{1--2 mm} & \colhead{2--3 mm} & \colhead{3--4 mm} & \colhead{4--6 mm} & \colhead{6--8 mm}  
& \colhead{8--10 mm} & \colhead{10--15 mm} \\
}
\startdata
10 & 0.06 & 0.05 & 0.05 & 0.07 & 0.09 & 0.10 & 0.14 & 0.13 \\
25 & 0.09 & 0.06 & 0.08 & 0.10 & 0.11 & 0.12 & 0.16 & 0.18 \\
\textbf{50} & \textbf{0.14} & \textbf{0.15} & \textbf{0.17} & \textbf{0.18} & \textbf{0.26} & \textbf{0.23} & \textbf{0.26} & \textbf{0.27} \\75 & 0.22 & 0.29 & 0.28 & 0.36 & 0.51 & 0.28 & 0.64 & 0.53 \\
90 &0.29 & 0.50 & 0.51 & 0.64 & 0.82 & 0.59 & 0.93 & 0.68 \\
 &  &  &  &  &  &  &  & \\
10 & 0.27 & 0.23 & 0.24 & 0.32 & 0.36 & 0.44 & 0.66 & 0.59 \\
25 & 0.43 & 0.30 & 0.35 & 0.44 & 0.53 & 0.51 & 0.73 & 0.78 \\
\textbf{50} & \textbf{0.60} & \textbf{0.56} & \textbf{0.65} & \textbf{0.77} & \textbf{1.02} & \textbf{0.79} & \textbf{1.10} & \textbf{1.01} \\
75 & 0.71 & 0.98 & 1.00 & 1.10 & 1.65 & 1.24 & 1.84 & 1.99 \\ 
90 & 0.92 & 1.52 & 1.48 & 1.86 & 2.40 & 2.23 & 2.55 & 2.22 \\
  &  &  &  &  &  &  &  & \\
Count & 37 &49 & 90 & 81 & 78 & 35 & 28 & 22 
\enddata
\tablecomments{The standard deviation (Rows 1--5) and peak-to-valley spread (Rows 6--10) of PWV measured in a night, binned by the mean PWV of that night. Based on the same set of data shown in Figure \ref{fig:apVisit_variation}. The median values are shown in boldface. All PWV values in millimeter.}
\end{deluxetable}

\end{document}